\RequirePackage{fix-cm}
\RequirePackage{lineno}
\documentclass[twocolumn]{svjour3}          
\usepackage[ruled,linesnumbered]{algorithm2e}
\usepackage{graphicx}
\usepackage{cuted}
\usepackage{microtype}

\usepackage{xspace}
\usepackage{hyperref}
\usepackage{listings}

\newcommand{\corsika}[1]{CORSIKA\xspace#1}
\newcommand{\iact}{\texttt{iact}\xspace}
\newcommand{\gcc}{GCC\xspace}
\begin{document}

\title{Optimizing Cherenkov photons generation and propagation in CORSIKA for CTA Monte--Carlo simulations}


\author{Luisa Arrabito \and Konrad Bernl{\"o}hr \and Johan Bregeon \and Matthieu Carr\`{e}re \and Adnane Khattabi \and Philippe Langlois  \and  David Parello  \and Guillaume Revy}

\institute{Luisa Arrabito \and Matthieu Carr\`{e}re \at
              Laboratoire Univers et Particules, Universit\'e de Montpellier Place Eug\`{e}ne Bataillon - CC 72, CNRS/IN2P3, F-34095 Montpellier, France \\
           \and 
           Konrad Bernl{\"o}hr \at
           Max-Planck-Institut f\"ur Kernphysik, P.O. Box 103980, D-69029 Heidelberg, Germany
           \and 
           Johan Bregeon \at
              Laboratoire Univers et Particules, Universit\'e de Montpellier Place Eug\`{e}ne Bataillon - CC 72, CNRS/IN2P3, F-34095 Montpellier, France \\
              now at Laboratoire de Physique Subatomique et de Cosmologie, CNRS/IN2P3, F-38026 Grenoble, France \\
           \and 
            Adnane Khattabi \at
              Universit\'e de Perpignan Via Domitia, Digits, Architectures et Logiciels Informatiques, F-66860 Perpignan, Universit\'e de Montpellier, Laboratoire d'€™Informatique Robotique, CNRS, France \\
              now at Laboratoire Bordelais de Recherche en Informatique, CNRS and Universit\'e de Bordeaux, F-33405 Talence, France
           \and Philippe Langlois \and David Parello \and Guillaume Revy \at
              Universit\'e de Perpignan Via Domitia, Digits, Architectures et Logiciels Informatiques, F-66860 Perpignan, Universit\'e de Montpellier, Laboratoire d'€™Informatique Robotique, CNRS, France 
}

\date{\today}
\maketitle
\newpage

\begin{abstract}
\corsika{} (COsmic Ray SImulations for KAscade) is a program for detailed simulation of extensive air showers initiated by high energy cosmic ray particles in the atmosphere, and is used today by almost all the major instruments that aim at measuring primary and secondary cosmic rays on the ground. The Cherenkov Telescope Array (CTA), currently under construction, is the next--generation instrument in the field of very--high--energy gamma--ray astronomy. Detailed \corsika{} Monte Carlo simulations will be regularly performed in parallel to CTA operations to estimate the instrument response functions, necessary to extract the physical properties of the cosmic sources from the measurements during data analysis. The estimated CPU time associated with these simulations is very high, of the order of 200 million HS06 hours per year. Code optimization becomes a necessity towards fast productions and limited costs. We propose in this paper multiple code transformations that aim to facilitate automatic vectorization done by the compiler, ensuring minimal external libraries requirement and high hardware portability.
\keywords{Gamma-ray astronomy \and Next generation Cherenkov telescopes \and HPC \and Code optimization \and Vectorization}
\end{abstract}

\section{Introduction}
\label{intro}
The Cherenkov Telescope Array (CTA) represents the next  generation  of ground--based instruments for very--high--energy gamma--ray astronomy \cite{CTA_2010bc}. CTA will operate almost 100 telescopes arranged in two arrays placed in each hemisphere: One on top of La Palma (Canary Islands, Spain) and one at Cerro Paranal (Chile). CTA is meant to exploit the imaging Cherenkov technique that consists in the detection and characterization of the Cherenkov light from Extensive Air Showers (EAS). When charged particles from EAS go faster than the speed of light in the atmosphere, blue light is emitted via the Cherenkov effect. High--energy gamma--rays from cosmic sources interact at the top of the atmosphere and generate an electromagnetic cascade that maps into an ellipsoid of blue light that is collected on ground by the array of telescopes. Each telescope is equipped with a high--speed and highly--sensitive camera that can take a short movie of the Cherenkov light flow. Eventually, information from all telescopes are put together to reconstruct the properties of the incident primary particle (nature, direction and energy), making it possible to do astronomy.

Cherenkov imaging instruments do rely heavily on full Monte--Carlo simulations at different stages, but especially to calibrate the event reconstruction, the selection of gamma--like events and to derive the full instrument response functions. CTA simulations rely on the \corsika software~\cite{Corsika1998} for the physics of EAS (including the generation of Cherenkov photons), on the \iact\texttt{/atmo} plugin for a fine handling of atmosphere properties and propagation of the Cherenkov light, and on \texttt{sim\_telarray}~\cite{2008APh....30..149B} for the ray tracing of light on the telescope mirrors and the response of the camera electronics. 

In recent years, very large simulation campaigns have been run by the CTA Consortium in order to finalize the design of the instrument and properly plan for the construction phase. In particular, both the site selection campaign~\cite{Hassan:2017paq} and the optimization of the array layout campaign~\cite{Acharyya:2019nwy} have run for around two years on thousands of computing nodes of the European computing grid (EGI), consuming more than 200 million HS06 \footnote{https://w3.hepix.org/benchmarking.html} CPU hours (or 2000 years of CPU time) and producing several Petabytes of data. 
In these simulation studies, about 70 \% of the CPU is required for the air shower simulation, about 30 \% for the telescope simulation and another small fraction for the last reconstruction and analysis steps. 

The optimization of the air shower simulation has hence become a major concern as it would bring flexibility in the management of large productions: we could either run productions in less time reducing the associated cost and the carbon footprint, or run larger productions for reduced statistical errors, for the same costs.

In this paper, we present our work for the optimization of the CTA simulation code, giving details both about the procedure we have followed and about the results obtained. In Section~\ref{sec:context}, we give a brief description of the simulation code and of the initial context of our project in order to explain better our choices. Then, in Section~\ref{code_analysis}, we examine the limitations of compilers when optimizing complex software and present the code profiling work that drove the choices made to develop our optimization procedure presented in Section~\ref{sec:opti} and that permitted us to track improvements at different levels. 
In particular, we propose code transformations to enable modern compilers to apply advanced optimizations which require suitably written code. Finally, we present the results of these optimizations in Section~\ref{perf}.

\section{Context}
\label{sec:context}
\corsika has been originally developed at FZKA \allowbreak Forschungszentrum Karlsruhe in the early 90s for air shower simulations for the KASCADE experiment. It was rapidly adapted by other collaborations for their use. The first were MACRO and HEGRA, then several other experiments across the world have been using \corsika for their simulations (Pierre Auger Observatory, IceCube, CTA, etc.).

The software consists of a main source, written in Fortran, handling the particle stack, the particle transport, the random number generation and the atmosphere description. 
Electromagnetic interactions simulation is also included in the main source using a tailored version of the EGS4~\cite{Nelson:1985ec} program.  Hadronic interactions models are implemented in different external packages used as plugins by the main program, also written in Fortran.

For imaging atmospheric Cherenkov experiments, like CTA, an additional package \iact\texttt{/atmo}~\cite{2008APh....30..149B}, written in C, handles the generation of Cherenkov photons within air showers, their propagation through the atmosphere as well as the 3D geometry of telescope arrays: See appendix~\ref{an:prop} for a schematic view of the Cherenkov photon production and propagation through the atmosphere. Telescopes are represented as fiducial spheres at different positions on the ground and only Cherenkov photons crossing these are saved in the final output. A more detailed description of the atmosphere is also supported in the \iact\texttt{/atmo} package allowing the use of external atmospheric profiles. 

Over 30 years, \corsika has evolved to become a large and hard to maintain complex scientific software, consisting of more than 10$^5$ lines of code. In this context, a project of a full rewriting of \corsika in modern C++ has started in 2018, here after named the \corsika 8 project~\cite{2018arXiv180808226E}~\cite{2019ICRC...36..236D}.
The project is led by KIT (Karlsruhe Institute for Technology), but is an effort of many groups from the whole air shower community (including the authors).
One of the main goals of \corsika 8 is to provide a flexible and modular framework to support physics applications, while maintaining a high computational speed.
Even if, in the long run, \corsika 8 will replace the current production version (\corsika 7), scientific collaborations like CTA will still rely on the latter for at least a few years, until \corsika 8 is fully validated. 
For this reason, we have decided to work on the optimization of \corsika 7, focusing on its use for CTA. The goal is twofold: First to obtain an improved version to be used already in the current-- and near--term CTA productions, second to identify the bottlenecks in the simulation process and test different optimization techniques that will be eventually transferred to \corsika 8.

There is not a unique approach to achieve faster simulations with \corsika, this depends very heavily on the use case, hence on the most important physics processes at stake and on the energy range. In the energy range of CTA (20 GeV -- 300 TeV), the simulation of a gamma--ray induced air shower (later called also \emph{event}) is relatively fast, taking typically from a few seconds to a few minutes for the highest energy showers.
However, high event statistics is needed to investigate shower fluctuations and the dependence on primary parameters.
Each production (defined by a given configuration of the instrument and the associated conditions) requires to simulate a few billions of showers, leading to an overall CPU time of about 60 million HS06 hours.
With shower events being independent from each other, the full production statistics can be obtained by running several \corsika jobs on different cores, each job simulating around 50 thousand showers in a few hours. 
By distributing the whole simulation workload over 8000 -- 10000 jobs running on the EGI infrastructure, the CTA Consortium is able to run a typical production in a period of about one month. 

For other experiments, like the Pierre Auger Observatory, the main issue is the very long CPU time needed to simulate single ultra--high energy showers (above 10$^{17}$ eV).  Indeed, the number of particles in an EAS, and hence the CPU time and disk space required for its full simulation, scales with the energy of the primary particle. As an example, a typical vertical shower induced by a hadron of 10$^{18}$ eV  requires a month of CPU time and about 100 Gigabytes of disk space.

Different specific methods have been introduced to speed--up the simulation of ultra--high--energy showers. These methods consist in introducing different kinds of approximations. Very early in the development of \corsika, so--called thinning algorithms were developed to reduce the number of particles by grouping them and assigning them weights to provide correct values of physical quantities on average. Another approach is to use the numerical solution of cascade equations instead of a full Monte--Carlo simulation. A complementary method, which does not affect the simulation results, consists in removing at an early stage of the simulation those particles that do not have any chance to be detected or which will not be useful.
Moreover, from a more computing oriented point of view, an alternative solution for ultra--high--energy simulations consists in parallelizing \corsika, i.e. distributing the single EAS simulation across many cores using the MPI protocol \cite{2015ICRC...34..528P}.

Yet from another perspective, the IceCube neutrino observatory, that measures high energy neutrinos from the cosmos, had to solve the difficult problem of propagating Cherenkov photons through the many layers of ice of the South pole. IceCube indeeds relies on \corsika for the simulation of air showers, but developed a dedicated photon propagation code. This part of the simulations was initially by far the most computation intensive part, and was successfully ported to GPUs~\cite{Chirkin:2013tma} to become a tiny fraction of the simulation time.

In the case of experiments like CTA, where the limitation comes from high event statistics rather than from extreme long events, there would be no gain in using the \corsika parallelized version. 
Our approach aims instead at reducing the CPU time of the sequential execution. 
To achieve this purpose, we have explored different techniques, like choosing appropriate compiler flags, code refactoring, algorithms change and vectorization, that we will discuss in Sections \ref{code_analysis} and \ref{sec:opti}. Our goal is to optimize the code with no impact on the accuracy of the results. 
For this reason, we did not consider techniques implying approximations, neither changing algorithms used for the implementation of physics processes. The only change at algorithmic level that we have introduced consists in an extended application of an already available algorithm for the fast interpolation of atmospheric profiles, as described in Section \ref{sec:atmoopt}. An additional constraint is that we want to be able to execute the optimized code on the EGI e--infrastructure that is currently used in production. 
Since the computing power of this e-infrastructure is essentially provided by CPUs, we did not consider off-loading part of the computations to GPUs. 
On the other hand, since EGI computing centers are equipped with CPUs of different generations, the advantage of vectorization is that the code can be easily ported on different types of CPUs.
Indeed, the principle of the vectorization is to make use of SIMD (Single Instruction on Multiple Data)~\cite{IEEETransComput} instructions to perform the same operation on multiple data simultaneously. These instructions are available on all the modern processors with different implementations, operating on registers of different sizes. The most common are: SSE4, AVX/AVX2, AVX--512, operating respectively on 128-bit, 256-bit and 512-bit registers. This means that a given operation can be executed on 4 doubles or 8 floats simultaneously for AVX/AVX2 and on 8 doubles or 16 floats for AVX--512.
The variation from an implementation to another being essentially the size of the registers, it is sufficient to parametrize the code at the compilation time to adapt it to the target CPU model.

\section{Code analysis and initial optimizations}
\label{code_analysis}

\subsection{Optimization compilation flags}
Before even starting to modify the code, we have extensively checked the compiler capabilities to automatically optimize the program.
This is done by choosing appropriate compilation flags that determine which optimization will the compiler attempt to do. 
However with this technique we have measured almost no performance improvement. The conclusion was that obtaining a meaningful performance gain requires modifying the code.

\subsection{Code Profiling}
In order to perform an effective code optimization, it is necessary to analyze the code and identify the potential bottlenecks that could be hindering the performances. For this purpose we have used the GNU profiler \textit{gprof}~\footnote{\href{https://sourceware.org/binutils/docs/gprof/}{https://sourceware.org/binutils/docs/gprof/}}~\cite{Graham:1982:GCG:872726.806987} together with the \textit{gprof2dot}~\footnote{\href{https://github.com/jrfonseca/gprof2dot}{https://github.com/jrfonseca/gprof2dot}} visualization add-on. This tool generates a call graph showing a detailed representation of the different functions of the code as well as the associated execution time. 
To obtain a representative profile we have executed \corsika with the same input parameters used in a standard CTA production for gamma--ray showers. We have adjusted the number of simulated showers to obtain an execution duration of about 10 minutes (2500 showers in the current case).

\begin{figure*}[htbp]
\centering
\includegraphics[width=\textwidth]{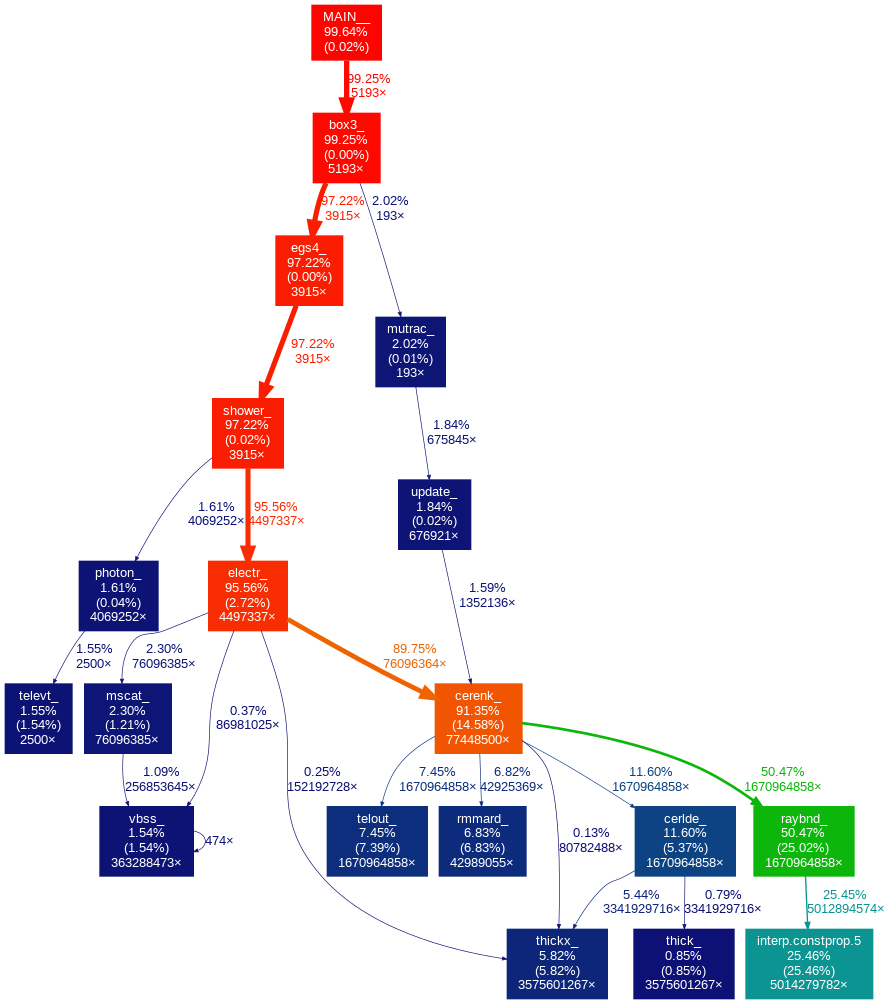}
\caption{Call graph generated with \textit{gprof} and \textit{gprof2dot}, for the profiling of \corsika (version 6.990 and \iact-\texttt{atmo} v1.51 compiled with \gcc v8.2.1) running 2500 gamma--ray showers in a standard CTA production configuration.}
\label{fig1}
\end{figure*}

In the call graph generated with \textit{gprof2dot}, the functions requiring a high computational cost are depicted in a red color palette and the colors shift to blues as the cost diminishes. Thus, we are able to observe the critical path of \corsika execution in Figure \ref{fig1}. Within this critical path, we observe that over 91 \% of CPU time is spent in the \texttt{CERENK} function which is responsible for the generation of Cherenkov photons within the air showers. The second most CPU intensive function in the critical path is the \texttt{RAYBND} function with over 50 \%. This function is responsible for the propagation of the photons in the atmosphere with refraction correction, meaning that \texttt{RAYBND} is called for every single photon bunch (over 1.7 billion for this execution, i.e. about 670 000 times per shower) which explains its high cost.

In order to cross-check these results we have performed the profiling with different tools, like Linux \emph{perf}, \emph{Callgrind} and a \emph{gdb}-based profiler, all giving compatible results \footnote{\href{https://gite.lirmm.fr/cta-optimization-group/cta-optimization-project/wikis/optimization_working_page}{https://gite.lirmm.fr/cta-optimization-group/cta-optimization-project/wikis}}. 
In particular, looking at the \emph{perf} profiling, which also includes low-level libraries, we observe that a large portion of CPU within \texttt{CERENK} and \texttt{RAYBND} is due to mathematical functions (\textit{exp}, \textit{asin}, \textit{cos}, \textit{sin}). 

Finally, we observe that in these functions the computations on the generated photons are intrinsically independent for each photon. From these observations, we can conclude that these functions present prime conditions for optimization through vectorization.

\subsection{Initial code optimization}
From the profiling results presented in the previous section, we expect that vectorizing \texttt{CERENK} and \texttt{RAYBND} is a very promising path for optimization. The vectorization of these two functions will be discussed in detail in Section \ref{sec:opti}.
In this Section, we discuss two other methods that we have employed to improve performances. The first consists in refactoring some parts of the code to reduce unnecessary computations, while the second consists in introducing slight changes at the algorithmic level.

\subsubsection{Code refactoring}
\label{sec:refactor}
Analyzing in more detail the code of the \texttt{RAYBND} function, we have realized that the atmospheric interpolation process, which accounts for one-fourth of the CPU spent in \texttt{RAYBND}, was not implemented in an optimal way.
The atmospheric interpolation concerns two types of tables: Atmospheric quality tables and refraction tables. 
Atmospheric quality tables contain the values of density, thickness and refraction index, while refraction tables contain coefficients for air refraction correction. The data points of all these tables correspond to different altitudes at non equidistant steps as more points are available around known altitudes showing greater variability of the atmosphere quality.
An interpolation algorithm is used to derive any of the 4 given variables at an intermediate altitude between two tabulated points. Typically, the used atmospheric tables consist of about 55 points.
The standard interpolation algorithm, implemented in the \texttt{RPOL} function, consists of two steps: First, a binary search algorithm retrieves the two closest points in altitude; Second a linear interpolation between these two points gives the value of the quantity of interest. 
Note that \texttt{RPOL} is not visible in the call graph of Figure~\ref{fig1} because of inlining. However, its child function (\texttt{INTERP}), performing the binary search, is shown and accounts for 25 \% of the CPU. 
In the \texttt{RAYBND} function, the \texttt{RPOL} function was called three times to calculate respectively the interpolated values of density, thickness and refraction index at the same altitude, meaning that exactly the same binary search was performed for each of the three calls. We have thus refactorized the code, so that the \texttt{RPOL} function performs the binary search only once for a given altitude, and then computes the interpolated values of the three variables. In this way, for each call to \texttt{RAYBND}, we have saved two binary searches. This simple modification resulted in a speed--up of 1.09 ($T_{ref}/T_{opt}$), which is already higher than what we have obtained using specific compiler flags.  

\subsubsection{Algorithmic change}
\label{sec:atmoopt}
Another optimization, already supported in \iact\texttt{/atmo}, consists in using a faster algorithm for atmospheric interpolation (called \emph{fast interpolation}). This algorithm builds fine--grained tables with uniform steps in altitude, using the same standard interpolation algorithm described above, but only once during an initialization phase. Interpolated values are then calculated from these fine--grained tables without the need of performing any binary search. On the one hand, these tables must be fine--grained enough so that the accuracy of looking up pre--defined values is similar to computing the interpolation on--demand, on the other hand the number of pre--calculated entries must be reasonable to limit the memory footprint. By running dedicated tests, we were able to show that using tables with 10000 pre--interpolated data points was good enough to keep an excellent agreement with a dedicated call to the linear interpolation routine.

In \iact\texttt{/atmo} v1.51, the \emph{fast interpolation} algorithm is implemented only for the atmospheric quality tables, but not for the refraction tables. Extending the usage of the \emph{fast interpolation} to all quantities was fully implemented in \iact\texttt{/atmo} v1.59, bringing to a speed--up of around 1.2.

\section{Optimization process}
\label{sec:opti}
Following the initial code analysis and clean--up, we were in good conditions to start the actual optimization and vectorization work on the main Cherenkov handling piece of code: The \texttt{CERENK} routine that produces Cherenkov photons and then calls the \texttt{RAYBND} routine to propagate photon bunches through the atmosphere.

\subsection{Preparing the code for vectorization}
Replacing scalar functions with their vectorized equivalents required some code transformations. We define \texttt{RAYBND\_VEC} as the vectorized version of \texttt{RAYBND} that processes multiple photon bunches together: All variables associated to photon bunches (\textit{e.g.} space and angular coordinates, arrival time, etc.) are replaced with vectors of specific length. 
The vector length can be adjusted through a compilation flag according to the available SIMD instructions on the processor and the corresponding size of the registers.
Since all these variables are in double precision, they can be stored into vectors of size 4 for AVX2 processors (256-bit registers) and of size 8 for AVX--512 processors (512-bit registers).
This transformation is necessary because the compiler is unable to vectorize entire functions and to apply all the changes necessary to the input parameters to switch from scalar to vector. The compiler is also unable to automatically replace the call to scalar \texttt{RAYBND} with the new vectorized version, meaning that we also had to unroll the main computation loop in \texttt{CERENK}, i.e. the loop over the particle sub-steps (see Appendix~\ref{an:prop}), to be able to call \texttt{RAYBND\_VEC}. 
Once this transformation is accomplished, it is also necessary to isolate the computations that define the parameters of these functions and extract them from the main loop so that the vectorization can be implemented.

\subsection{Vectorization of mathematical functions}
As shown in the previous section, an important portion of the computational needs of \texttt{RAYBND} comes from mathematical functions. \corsika originally employs implementations of these functions from the linux \textit{libm} library to do computations. A number of vectorized mathematical libraries are available on the market, among which Intel SVML~\footnote{\href{https://software.intel.com/en-us/cpp-compiler-developer-guide-and-reference-overview-intrinsics-for-short-vector-math-library-svml-functions}{https://software.intel.com/en-us/cpp-compiler-developer-guide-and-reference-overview-intrinsics-for-short-vector-math-library-svml-functions}}, AMD libm~\footnote{\href{https://developer.amd.com/amd-aocl/amd-math-library-libm/}{https://developer.amd.com/amd-aocl/amd-math-library-libm/}}, the CERN VDT~\footnote{\href{https://github.com/dpiparo/vdt}{https://github.com/dpiparo/vdt}}~\cite{2014JPhCS.513e2027P} and the \textit{SIMD vector libm}~\cite{lauter:hal-01511131}. Intel SVML and AMD \textit{libm} being proprietary, we tested only VDT and the \textit{SIMD vector libm} that we found to show similar performances.  We decided to use the \textit{SIMD vector libm} because it is based on the principle of automatic vectorization, meaning that the functions are implemented such that the compiler is able to vectorize them based on the architecture of the machine the code is built on, ensuring the portability of the library and its stand--alone aspect.

\subsection{Instruction vectorization}
With the vectorization of mathematical functions accomplished, the next step in vectorization--based optimizations is to attempt the vectorization of single instructions. Portability being a high constraint, we aim to limit external libraries dependencies to the maximum and focus on automatic vectorization. We will discuss in this section the limitations the compiler faces during the process of automatic vectorization and possible ways to bypass them.

\subsubsection{Restructuring tests}
The basis for automatic vectorization consists of enabling the usage of SIMD instructions with the appropriate compilation flags. By simply doing so for \corsika in its original code structure, we observe in the assembler code, that the compiler manages to vectorize very few instructions, mostly due to the complexity of the code. Our optimization target for vectorization, the \texttt{RAYBND} function, often exhibits code parcels with a structure represented in Algorithm~\ref{algo1}. We usually have multiple nested conditional statements that hide the most common computation (line 9 in Algorithm~\ref{algo1}). In fact, by testing the occurrences in which these conditions are validated, we observe that they are very rare and that they require simulations of a large number of showers before manifesting. Therefore, eliminating the hindrance to automatic vectorization requires restructuring the code to extract the main computation from inside these conditional statements. 
\begin{algorithm}[htb]\label{algo1}
\SetAlgoLined
\SetKwProg{Fn}{Function}{}{end}
\Fn{main()}{
\For{($ i = 0$; $i < n\_bunches$; $i$++)}{
\eIf{Condition1}{
  Compute case 1; \\
  }{
  \eIf{Condition2}{
   Compute case 2;  \\
     }{
   Compute general case;  \\
     }
   }
}
}
\caption{Original code structure.}
\end{algorithm}

\begin{algorithm}[htb]\label{algo2}
\SetAlgoLined
\SetKwProg{Fn}{Function}{}{end}
\Fn{main()}{
Test1 = False;\\
Test2 = False;\\
\For{($ i = 0$; $i < n\_bunches$; $i$++)}{
  Test1 = Test1 OR Condition1(i);\\
  Test2 = Test2 OR Condition2(i);\\
}
\If{Test1 OR Test2}{ 
  Scalar computation using Algorithm \ref{algo1} ;
  Return;
}
\For{($ i = 0$; $i < n\_bunches$; $i$++)}{
 Compute general case;
}   
}
\caption{Restructured code for vectorization.}
\end{algorithm}

The first step for this transformation is to define a vector of Cherenkov photon bunches in order to call the vectorized function \texttt{RAYBND\_VEC}: This has been done in the preparation step for the vectorization of mathematical functions. The \texttt{CERENK} routine has also been unrolled and we are now able to define two execution scenarios. 
The first scenario, the most common one, consists in computing the general case isolated from the rest of the calculations inside a loop of the size of the vector of photon bunches ($n\_bunches$).
In this way the compiler can easily identify it as vectorizable (line 12 in Algorithm~\ref{algo2}). In the second scenario, that happens a lot less frequently, all conditions in the \textbf{if} statements in Algorithm~\ref{algo1} are gathered and tested first for every photon bunch in the vector (lines 5 and 6 in Algorithm \ref{algo2}). If one single occurrence of these conditions is verified, we immediately switch to the scalar execution of Algorithm~\ref{algo1} and return. A sequence of transformations of this type were used to build \texttt{RAYBND\_VEC}: We start by concatenating all test scenarios and executing scalar computations if necessary; followed with a series of small easily vectorizable loops containing just a few lines of code.

Appendix~\ref{an:sample} proposes a parallel view of the original code and the optimized code in order to give the reader a better understanding of the implementation and an idea of the readability of the optimized code.

\subsubsection{Calls to function}
We are able to increase the portion of vectorizable code in \texttt{RAYBND} thanks to the isolation of instructions and restructuring of tests presented in the previous section. These transformations also enable an easier analysis of the code in order to understand any additional reasons for which the compiler would be unable to vectorize some parts of the code. This analysis was achieved by tracing the compiler behavior and inspecting the assembly code. We found out that the compiler is not allowed to automatically vectorize the code when a function call is present within an instruction. In the case of \texttt{RAYBND}, most functions calls of this nature are used for the interpolation process on atmosphere parameters (see Section~\ref{sec:refactor}). We also observe that these calls are often repeated for multiple instructions with the same input values. As shown in Algorithm~\ref{algo3}, when a secondary function \texttt{Interpolation} is hidden within the call to \texttt{MainInterpolation}, the compiler is unable to detect that the calls are identical with the parameter \texttt{Q}. In most cases, the compiler is able to detect redundant function calls and avoid these repetitions. However, because the code is designed with many small functions and shows multiple indirect function calls, the compiler often fails to optimize large parts of the code. Refactoring was done to manually extract these function calls from their respective instructions and define variables at the beginning of the function that contain the return values of the mentioned function calls. With these values at hand, we are able to place the variables instead of the function calls in the instructions so that the compiler successfully manages to further vectorize the code. One should note that this step would not have been necessary if indirect function calls were not present as a result of over splitting the code into small functions.
\begin{algorithm}[htb]\label{algo3}
  \SetAlgoLined
  \SetKwProg{Fn}{Function}{}{end}
  \Fn{MainInterpolation(P, Q)}{
  ...\\
  Interpolation(Q);
  }
  \Fn{main()}{
    MainInterpolation($P_1$, Q);\\
    MainInterpolation($P_2$, Q);\\
    MainInterpolation($P_3$, Q);\\
  } 
  \caption{Indirect repeated function calls.}
\end{algorithm}

\subsubsection{Code translation}
Optimizations discussed so far were introduced to \corsika in the \texttt{RAYBND} function, first because it is the second most CPU time consuming function, and additionally because it is coded in C, a language for which many tools and libraries exist to study and implement vectorization. However, the main routine, \texttt{CERENK}, is written in Fortran. By unrolling the main computation loop in \texttt{CERENK} and enabling the vectorization flag during the compilation, we observe very little vectorization of the code. Furthermore, with the intentions of replicating the same optimization techniques we employed on \texttt{RAYBND}, we decided to translate \texttt{CERENK} from Fortran to C. Multiple difficulties arise from this switch of programming languages. First, most of the data structures in the Fortran portion of \corsika, are defined in common blocks which represent global variables that all functions are able to access. Translating the function into C meant that we had to define these common blocks in C-appropriate data structures while maintaining the same format and variable order as the common blocks in Fortran. \texttt{CERENK} also includes some highly complex conditional statements that required some attention in making sure the C equivalent depicted the same execution scenarios. These cases being a rare occurrence, a wide range of tests were deployed to verify the validity of the C version of \texttt{CERENK} in all possible scenarios. Finally, \corsika contains multiple packages and thus different configuration scripts. Therefore, linking the new \texttt{CERENK} function for seamless integration into the general compilation process of \corsika required a restructuring of the way the program is packaged in order to avoid disturbing other portions of the code.

After translating \texttt{CERENK} to C, we are able to apply the optimization techniques we previously used for \texttt{RAYBND}. The restructuring of tests for \texttt{CERENK} is more complicated than \texttt{RAYBND} with few cases where it is impossible to isolate an instruction from the conditional statement. These constraints could be overcome with some algorithmic changes to the code but for the sake of maintaining the  overall structure of the code, we chose to retain some \textbf{if} expressions in loops that would be vectorizable otherwise.

In addition, \corsika allows to optionally store the longitudinal development of Cherenkov photons production that should match the altitude profile of the atmospheric density. This profile is computed in a function named \texttt{CERLDE}, that is called inside \texttt{CERENK} . However, we are unable to use the original \texttt{CERLDE} function with the vectorized versions of \texttt{RAYBND} and \texttt{CERENK} since the computations done in \texttt{CERLDE} are scalar and use data structures that are not adequate for the vectorized functions. Therefore, we had to apply the same transformations as 
\texttt{CERENK}: Translating to C and optimizing for automatic vectorization. We keep the scalar version of \texttt{CERLDE} for the computation associated to the photons that remain after the distribution into vectors.

\section{Performance analysis}
\label{perf}
In this section, the gains in performance obtained after the various steps of optimization described in Section \ref{sec:opti}, will be presented together with their impact on numerical results.

\subsection{Experimental setup}
\subsubsection{Code versions}
Starting with a reference version of the code, named "V-ref", corresponding to the original source code distributed by the developers, we were able to define a coherent sequence of optimization steps. The first step and main bottleneck is the \texttt{RAYBND} routine, that was actually vectorized in two phases. The version named "V--ray--0" is our initial attempt including code reorganization and vectorization of critical computations, and is the one presented in~\cite{2019EPJWC.21405041A}. The version named "V--ray" corresponds to the full code restructuring and vectorization of all mathematical and single instruction calls. The second step is the translation of \texttt{CERENK} to C and its optimization through the full vectorizaton process in order to obtain the most advanced vectorized versions of both \texttt{RAYBND} and \texttt{CERENK}, that are named "V--cer".

\begin{figure}[htbp]
\centering
\includegraphics[width=\linewidth]{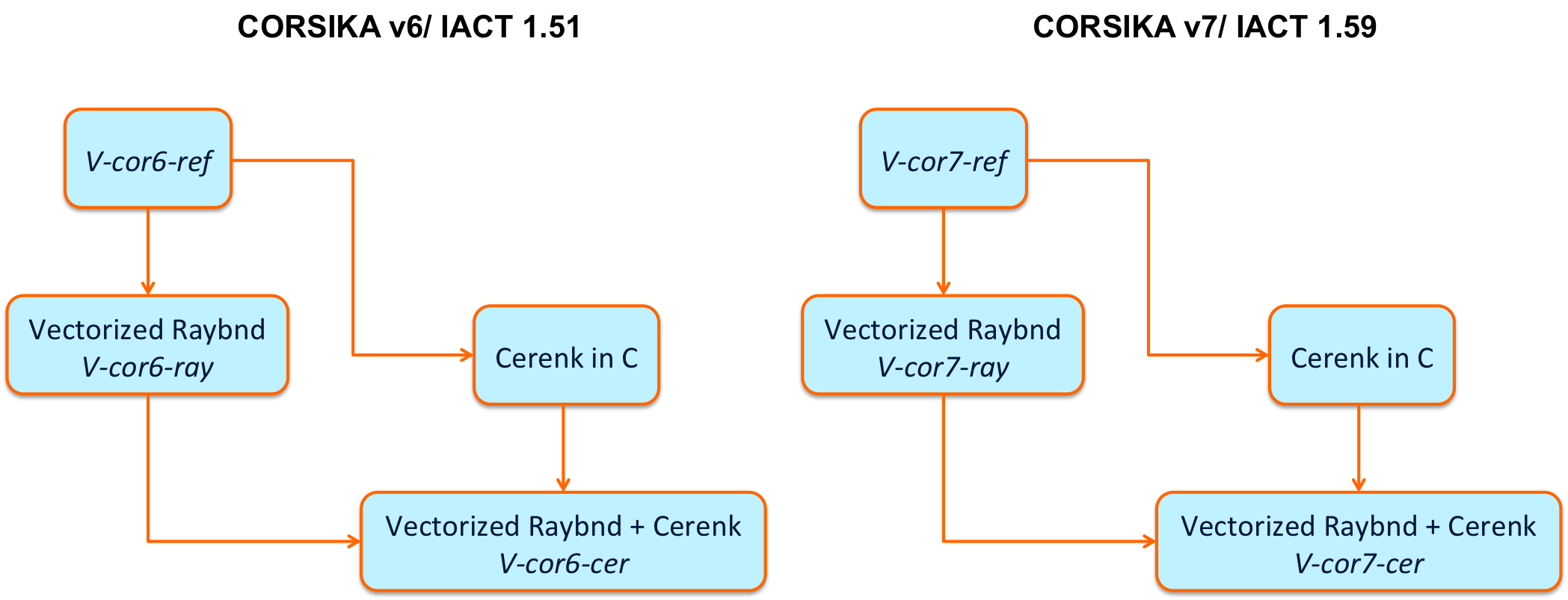}
\caption{Optimized versions of CORSIKA 6 and 7.}
\label{fig2}
\end{figure}

This optimization sequence was first designed and applied on CORSIKA version 6.990 with the \iact-\texttt{atmo} package v1.51, and then also applied on CORSIKA version 7.69 with \iact-\texttt{atmo} v1.59 that is the most recent version planned to be used for the next CTA large productions. The different optimized versions of CORSIKA are schematically presented on Figure~\ref{fig2}.

\subsubsection{Test environment}
\label{sec:testenv}
The different versions of \corsika were deployed on a dedicated server running  CentOS 7.5.1804. The processor is an Intel Xeon Gold x86\_64 with 16 cores running at 3.60 GHz. The memory cache is distributed on three levels: 32 KB in L1, 1024 KB in L2 and 16896 KB in L3. The processor has both AVX2 and AVX--512 advanced vector extensions available. We relied mostly on the \gcc compiler v8.2.1, but also verified that the code compiled and run with older versions starting from \gcc v4.8.7, that is the default on CentOS 7.5. All the performance results that follow correspond to code compiled with \gcc v8.2.1 and compilation flags \texttt{-O3} \texttt{-mavx2}. 

The numerical validation of the computations is based on a python script that verifies that the \corsika output file containing the Cherenkov photons position, angle and arrival time parameters, is bit-wise identical to the output obtained when executing the reference version. 

The \textit{perf} tool~\footnote{\href{https://perf.wiki.kernel.org/index.php/Main_Page}{https://perf.wiki.kernel.org/index.php/Main\_Page}} was used to measure the CPU time with a non intrusive sampling that counts the cycles and instructions during the execution: \texttt{perf stat -e cycles,instructions}. For each \corsika version, we have considered the average CPU time over 10 measurements, for which we observed fluctuations less than 1\%.

\subsection{Performance gain with AVX2 instructions}
Performances have been assessed for two execution scenarios. 
The first one is a testing configuration of \corsika using CTA standard productions parameters, with the exception that it does not include the calculation of the Cherenkov photons longitudinal profile, performed by the \texttt{CERLDE} function.
The second scenario corresponds to the final \corsika{7} configuration as used in CTA standard productions, which also required the \texttt{CERLDE} vectorization.

To achieve an accurate measurement of the performances, we have configured \corsika to simulate a representative sample of 2500 gamma--ray induced air showers. 
An additional configuration option was used to ensure that exactly the same sequence of random numbers was used for all physics processes. With this configuration, we made sure that we were able to run exactly the "same" 2500 showers through all processes so that computation time measurement corresponded to exactly the same set of operations.

\subsubsection{CORSIKA 6 and 7 performances for a testing configuration}
\begin{figure}[htbp]
	\centering
	\includegraphics[scale = 0.47]{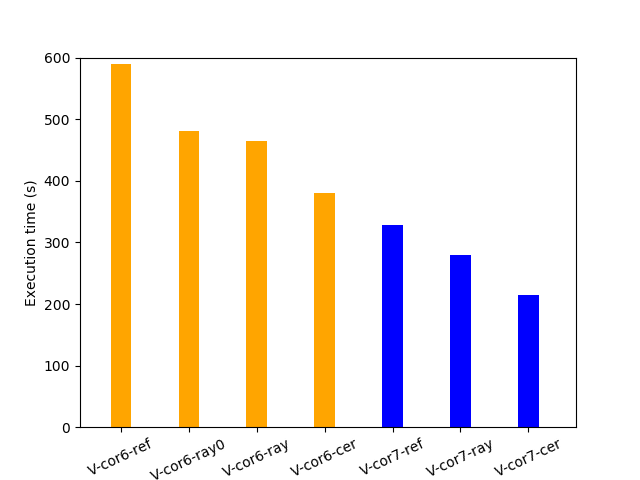}
          \caption{CPU time measurements of the different \corsika{} versions for a testing configuration, which does not include the calculation of Cherenkov photons longitudinal profiles. Computations are in double precision with AVX2 instructions enabled. The reported values correspond to the average over 10 measurements for 2500 shower runs.}
	\label{fig:cputime}
\end{figure}

Looking first at the execution time for the different optimized versions of \corsika{6}, presented as orange bars on the graph of Figure~\ref{fig:cputime}, we observe that with the initial work done on the code rationalization, and the vectorization of the most important calls to mathematical functions, so going from "V-cor6-ref" to "V-cor6-ray0", we were already able to speed--up the execution time by a factor 1.23. Including the full vectorization of \texttt{RAYBND}, we could gain a bit more with a speed--up of 1.27 for "V-cor6-ray". From there, Fortran code translation to C of \texttt{CERENK}, and significant refactoring to expose automatic vectorization were necessary to reach the final speed--up of 1.55 measured on "V-cor6-cer". The same optimization process applied on \corsika{7}, lead us from "V-cor7-ref" to "V-cor7-ray" with a speed--up factor of 1.18, and then to "V-cor7-cer" with a speed--up of 1.53.

Moreover, it's important to note that the important drop in execution time observed when comparing "V-cor6-ref" and "V-cor7-ref" is due to the fact that the new interpolation scheme (see Section~\ref{sec:atmoopt}) was introduced in the \iact--\texttt{atmo} v1.59 that is used with the \corsika{7} package. This new scheme also means that the code spends significantly less time in \texttt{RAYBND} in \corsika{7} than what did in \corsika{6}, meaning that optimizing \texttt{RAYBND} alone seems less effective than it did with the previous version, corresponding to  speed up of 1.27 for \corsika{6} as opposed to 1.18 for \corsika{7}.

\subsubsection{CORSIKA 7 performances for a CTA standard configuration}
\begin{figure}[htbp]
	\centering
	\includegraphics[scale = 0.47]{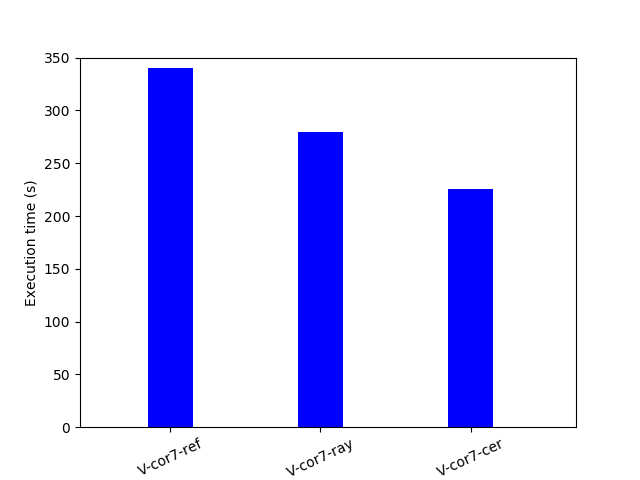}
	\caption{CPU time measurements of \corsika{7} (reference and optimized versions) for the final configuration as in CTA productions. Computations are in double precision with AVX2 instructions enabled. The reported values correspond to the average over 10 measurements for 2500 shower runs.}
	\label{fig:cputime_cerlde}
\end{figure}

When CTA simulations are run for large productions, \corsika is configured to save the longitudinal profile of Cherenkov photons, calculated in the \texttt{CERLDE} routine. \texttt{CERLDE} was translated to C and fully optimized only for \corsika{7}, and performance measurements are presented on the bar graph of Figure~\ref{fig:cputime_cerlde}. The full optimization of \texttt{RAYBND} lead us from "V-cor7-ref" to "V-cor7-ray" and a speed--up factor of 1.22. We then manage to reach a speed--up of 1.51 for a final optimized version of \corsika{7} ("V-cor7-cer") with all three functions \texttt{RAYBND}, \texttt{CERENK} and \texttt{CERLDE} vectorized. 

\section{Conclusion and prospects}
In this work, we have focused on the automatic vectorization of the \corsika air shower simulation program for the use case of CTA. An analysis of the code using different profiling tools has shown that the generation of Cherenkov photons (\texttt{CERENK} routine) and their propagation in the atmosphere (\texttt{RAYBND} routine) are the heaviest steps of the simulation CPU time wise. We have defined an optimization process applied to these two steps that aims at facilitating the automatic vectorization of the code by the compiler. The foremost important action was to clean--up and rationalize the code, which already lead to significant improvements without even starting vectorization. Then, we were able to restructure the tests to extract the main computational instructions from inside conditional statements and to make sure that function calls were called from outside these instructions. We also relied on the \textit{SIMD vector libm}, a library that implements the automatic vectorization of mathematical functions. 
The optimized version is integrated in the official \corsika \iact\texttt{/atmo} package used in CTA and can be simply activated by a specific compilation flag.
In Appendix~\ref{an:profile} we report the profile of \corsika after the applications of all these transformations.  
With these optimizations, we are able to reach a speed--up of 1.51 for \corsika{7} with AVX2 instructions, considering a configuration used for large--scale CTA simulation productions.
It's worth mentioning that the actual improvement in simulation efficiency since the beginning of our work on \corsika{6} reaches almost a factor of 2: This translates in significantly reduced costs and carbon footprint, or more resources available to achieve higher number of event statistics when needed.

We plan further work on the optimization of \corsika{7}, starting with an in--depth study of the accuracy of the computations of the most CPU intensive steps. If the precision of such steps can be reduced from double to float, vectorization will be made twice as more efficient. 
A second path forward is to analyze how the cache memory is accessed in order to understand if some computations may be reorganized to manage memory in a more efficient way, and further minimize the overall CPU time.

From the case study of the optimization of \corsika, we were able to devise and demonstrate a general transformation process of high--performance code for automatic vectorization conditioning. The process starts with the extraction of the most common instructions to place them in the fore--front of the code to ensure the compiler's capacity to vectorize the primary computations. This transformation requires mostly restructuring the conditional expressions as well as dividing long complex loops into smaller more vectorizable loops. Then, we observed that excessive splitting of the code into functions can lead into having multiple layers of indirect function calls that hinder the compiler from automatically replacing redundant function calls which in turn prevents instruction--level vectorization when these instructions contain a function call.

Based on this optimization methodology, we envision that at least part of this process could be handled automatically through a high--level tool that would be able to analyze the code and propose well targeted optimizations without any prior knowledge on the details of the code. We also imagine such a tool being extended into a fully automatic optimizer of HPC applications that could prove to be extremely challenging for manual optimization.

\section*{Acknowledgements}
This project has received financial support from the CNRS through the MITI interdisciplinary programs. 
We thank our colleagues from the CORSIKA simulation package, in particular Tanguy Pierog and Ralf Ulrich, for many fruitful discussions.
We are also looking forward to continue our work on optimizing air shower simulations within the CORSIKA 8 project.

\bibliographystyle{spphys}       
\bibliography{refs}   

\begin{thebibliography}{10}
\providecommand{\url}[1]{{#1}}
\providecommand{\urlprefix}{URL }
\expandafter\ifx\csname urlstyle\endcsname\relax
  \providecommand{\doi}[1]{DOI \discretionary{}{}{}#1}\else
  \providecommand{\doi}{DOI \discretionary{}{}{}\begingroup
  \urlstyle{rm}\Url}\fi

\bibitem{CTA_2010bc}
M.~Actis, et~al., Exper. Astron. \textbf{32}, 193 (2011).
\newblock \doi{10.1007/s10686-011-9247-0}

\bibitem{Corsika1998}
D.~Heck, J.~Knapp, J.~Capdevielle, G.~Schatz, T.~Thouw, Corsika: A monte carlo
  code to simulate extensive air showers.
\newblock Tech. rep. (1998).
\newblock \doi{10.5445/IR/270043064}.
\newblock \urlprefix\url{http://bibliothek.fzk.de/zb/abstracts/6019.htm}.
\newblock 51.02.03; LK 01; Wissenschaftliche Berichte, FZKA-6019 (Februar 98)

\bibitem{2008APh....30..149B}
K.~{Bernl{\"o}hr}, Astroparticle Physics \textbf{30}(3), 149 (2008).
\newblock \doi{10.1016/j.astropartphys.2008.07.009}

\bibitem{Hassan:2017paq}
T.~Hassan, et~al., Astropart. Phys. \textbf{93}, 76 (2017).
\newblock \doi{10.1016/j.astropartphys.2017.05.001}

\bibitem{Acharyya:2019nwy}
A.~Acharyya, et~al., Astropart. Phys. \textbf{111}, 35 (2019).
\newblock \doi{10.1016/j.astropartphys.2019.04.001}

\bibitem{Nelson:1985ec}
W.R. Nelson, H.~Hirayama, D.W. Rogers, {The Egs4 Code System}.
\newblock Tech. rep., SLAC (1985)

\bibitem{2018arXiv180808226E}
R.~{Engel}, D.~{Heck}, T.~{Huege}, T.~{Pierog}, M.~{Reininghaus}, F.~{Riehn},
  R.~{Ulrich}, M.~{Unger}, D.~{Veberi{\v{c}}}, arXiv e-prints arXiv:1808.08226
  (2018)

\bibitem{2019ICRC...36..236D}
H.~{Dembinski}, L.~{Nellen}, M.~{Reininghaus}, R.~{Ulrich}, in \emph{36th
  International Cosmic Ray Conference (ICRC2019)} (2019), p. 236

\bibitem{2015ICRC...34..528P}
T.~{Pierog}, R.~{Engel}, D.~{Heck}, G.~{Poghosyan}, J.~{Oehlschl{\"a}ger},
  D.~{Veberic}, in \emph{34th International Cosmic Ray Conference (ICRC2015)}
  (2015), p. 528

\bibitem{Chirkin:2013tma}
D.~Chirkin, Nucl. Instrum. Meth. \textbf{A725}, 141 (2013).
\newblock \doi{10.1016/j.nima.2012.11.170}

\bibitem{IEEETransComput}
M.~{Flynn}, IEEE Transactions on Computers \textbf{C-21}, 948 (1972).
\newblock \doi{10.1109/TC.1972.5009071}

\bibitem{Graham:1982:GCG:872726.806987}
S.L. Graham, P.B. Kessler, M.K. Mckusick, SIGPLAN Not. \textbf{17}(6), 120
  (1982).
\newblock \doi{10.1145/872726.806987}.
\newblock \urlprefix\url{http://doi.acm.org/10.1145/872726.806987}

\bibitem{2014JPhCS.513e2027P}
D.~{Piparo}, V.~{Innocente}, T.~{Hauth}, in \emph{Journal of Physics Conference
  Series v.513} (2014), p. 052027.
\newblock \doi{10.1088/1742-6596/513/5/052027}

\bibitem{lauter:hal-01511131}
C.~Lauter, in \emph{{2016 50th Asilomar Conference on Signals, Systems and
  Computers }} (Pacific Grove, United States, 2016), pp. 407 -- 411.
\newblock \doi{10.1109/ACSSC.2016.7869070}.
\newblock \urlprefix\url{https://hal.archives-ouvertes.fr/hal-01511131}

\bibitem{2019EPJWC.21405041A}
L.~{Arrabito}, et~al., in \emph{European Physical Journal Web of Conferences
  v.214} (2019), p. 05041.
\newblock \doi{10.1051/epjconf/201921405041}

\end{thebibliography}

\appendix
\onecolumn
\section{Simulation of the Cherenkov photon production and propagation}
\label{an:prop}
\begin{figure*}[htbp]
	\centering
        \includegraphics[width=0.9\textwidth]{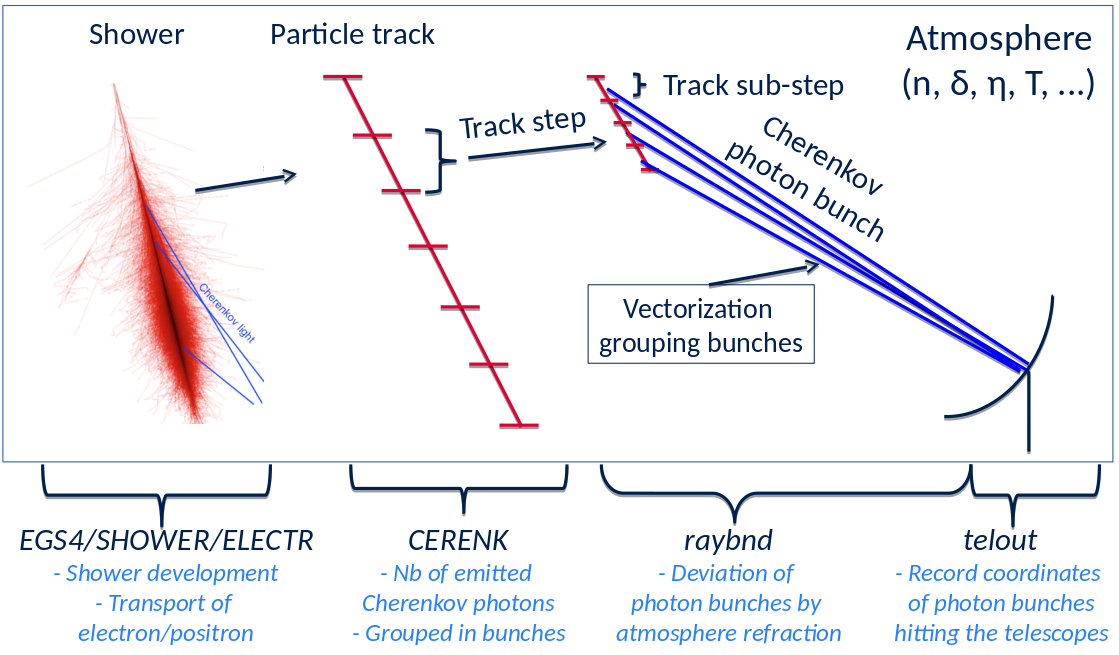}
	\caption{Schematic view of the code flow from the electromagnetic shower to the production and propagation of the Cherenkov photons.}
	\label{fig:cerenk}
\end{figure*}

Figure~\ref{fig:cerenk} presents a schematic view of the simulation steps of a typical electromagnetic event with \corsika.
Shower development and the electron/positron transport are  handled in the \texttt{EGS4}, \texttt{SHOWER} and \texttt{ELECTR} subroutines.
Then, each particle track in the shower is subdivided into several steps and for each step, the number of emitted Cherenkov photons is calculated by the \texttt{CERENK} subroutine. In order to reduce the computing time of photon propagation, all the computations are applied to bunches of typically 5 photons rather than to individual photons. Particle steps are further subdivided into sub-steps so that a single photon bunch is emitted at each sub-step. At each sub-step iteration the \texttt{RAYBND} function is called to calculate the bending of the photon bunch due to the refraction in the atmosphere and its propagation toward the ground. Finally, the coordinates of the photon bunches intersecting the telescope geometry are recorded and saved in the \corsika output (\texttt{TELOUT} function).

\newpage

\section{Sample of codes pre-- et post--optimization}
\label{an:sample}
To give the reader an idea of the optimized code looks like, two code listings are following, left being the original version, and right the optimized version that is automatically vectorized by \gcc. \texttt{VECTOR\_SIZE} is set according to the length of the SIMD register available on the CPU.

\noindent
\begin{minipage}{0.49\linewidth}
\begin{lstlisting}[title=Original]
// Emission direction replaced
// by observed direction.
*u *= sin_t_obs/sin_t_em;


*v *= sin_t_obs/sin_t_em;


// Downward ray remains downward ray
if ( (*w) >= 0. )
*w = sqrt(1.-sin_t_obs*sin_t_obs);
// Upward ray remainy upward
else 
*w = -sqrt(1.-sin_t_obs*sin_t_obs); 

// Position in observation level
// corrected for displacement. 
*dx += hori_off * (*u)/sin_t_obs;


*dy += hori_off * (*v)/sin_t_obs;



// Light travel time added to emission time.
*dt += travel_time;


\end{lstlisting}
\end{minipage}\hfill
\begin{minipage}{0.49\linewidth}
\begin{lstlisting}[title=Vectorized]
for(int i=0; i< VECTOR_SIZE; i++){
u[i] *= sin_t_obs[i]/sin_t_em[i];
}

for(int i=0; i< VECTOR_SIZE; i++){
v[i] *= sin_t_obs[i]/sin_t_em[i];
}

for(int i=0; i< VECTOR_SIZE; i++){
if(w[i] >= 0) 
w[i] = sqrt(1.-sin_t_obs[i]*sin_t_obs[i]);
else          
w[i] = -sqrt(1.-sin_t_obs[i]*sin_t_obs[i]);
}

for(int i=0; i< VECTOR_SIZE; i++){
dx[i] += hori_off[i] * (u[i])/sin_t_obs[i];
}

for(int i=0; i< VECTOR_SIZE; i++){
dy[i] += hori_off[i] * (v[i])/sin_t_obs[i];
}

for(int i=0; i< VECTOR_SIZE; i++){
dt[i] += travel_time[i];
}
\end{lstlisting}
\end{minipage}

\newpage

\section{CORSIKA profiling after optimization}
\label{an:profile}
\begin{figure*}[htbp]
	\centering
	\includegraphics[width=0.9\textwidth]{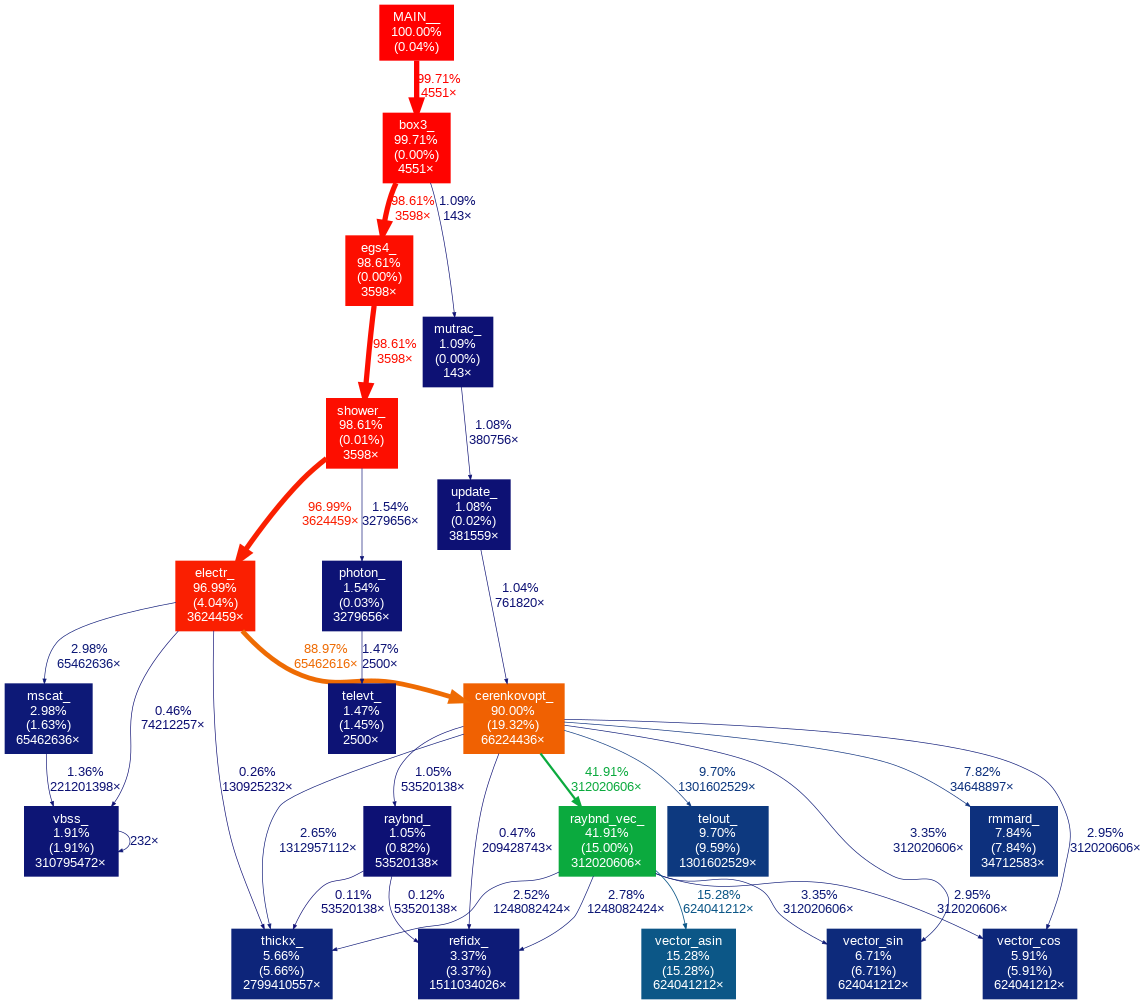}
	\caption{Profile of the \corsika optimized version for an execution with 2500 gamma--ray showers on an AVX2-enabled processor.}
	\label{fig:profile}
\end{figure*}

Figure~\ref{fig:profile} shows the profiling of \corsika after all the optimizations described in Sections \ref{code_analysis} and \ref{sec:opti}.
The critical execution path is the same as for the original version (see Figure \ref{fig1}), where \texttt{CERENK} and \texttt{RAYBND} have been replaced by their vectorized counterparts (\texttt{CERENKOVOPT} and \texttt{RAYBND\_VEC} respectively). 
The effect of the vectorization is visible in the reduction of the number of calls to \texttt{RAYBND\_VEC} with respect to \texttt{RAYBND} in Figure \ref{fig1}. It should also be noticed that the original scalar version of \texttt{RAYBND} is still called as fall back in about 1 \% of cases to treat the photon bunches that do not fit into 4-lenght vectors. Finally, the extended usage of vectorized mathematical functions in \texttt{CERENKOVOPT} and \texttt{RAYBND\_VEC} is shown in the call graph. 

\end{document}